\documentstyle[preprint,aps]{revtex}

\begin{document}

\begin{titlepage}
\begin{flushright}
NSF-ITP-95-138\\
NUB-TH-3129\\
\end{flushright}

\begin{center}
\Large{\bf Hierarchies and Textures in Supergravity Unification}\\
Pran Nath \\
\small{\it Institute for Theoretical Physics \\
University of California \\
Santa Barbara, CA 93106-4030 \\
\& \\
Department of Physics, Northeastern University \\
Boston, MA 02115}\footnote{Permanent address} \\
\date{\today}
\end{center}
\begin{abstract}
It is proposed that supergravity unified models contain in addition to the
hidden and the visible sectors, a third sector which contains exotic matter
with couplings to the fields of both the visible and the hidden sectors.
After spontaneous breaking of supersymmetry exotic matter becomes superheavy
and its elimination leads to quark-lepton textures at the GUT scale with
a hierarchy in powers of $M_G/M_P$.
Textures in the Higgs triplet sector
are also computed and it is shown that
 they modify predictions of
proton decay in unified models.
\end{abstract}
\end{titlepage}

One of the puzzles of particle physics is the question of how the quark-lepton
mass hierarchies arise and attempts have been made over the years to resolve
this problem \cite{aref}.
An interesting observation towards a solution
 is by Georgi and Jarlskog \cite{ref1} that the quark-lepton mass pattern
can be
understood in terms of textures at the GUT scale, and there have been
many further attempts to implement and explore this idea[3-7].
The simplest GUT theories,the ordinary or the SUSY ones, do not
contain such textures. Thus while the minimal supersymmetric SU(5) model
        predicts $m_b/m_{\tau}$ which is in fair agreement with experiment
the ratios $m_s/m_{\mu}$ and $m_d/m_e$ are off by an order of magnitude.
In this Letter we present a supergravity framework which incorporates
the
textures. Our starting point is the assumption that there exists a new kind
of
matter (exotic matter)  beyond the spectrum of the minimal supergravity model
 and that this matter
couples to both the visible sector fields and to the fields of the hidden
sector \cite{ref10,ref3}.
The above assumption leads us to include a third sector in
supergravity unified models where such matter resides[10].
After spontaneus breaking of supersymmetry the exotic matter become superheavy
with masses of size O($M_P$), where $M_P$ is an effective superheavy scale
of order the Planck scale.
Integration over the spectrum of the exotic matter
leads to a set of Planck slop terms scaled by
powers of $M_G/M_P$ at the GUT scale.
We then  show that Planck slop terms of this type are sufficient to generate,
for example,
the full Georgi-Jarlskog textures. We also compute in this model
the textures in the Higgs triplet sector and find that they are different
from those in the quark lepton mass sector. This result has important
implications for proton decay lifetimes which we also discuss.

We give now the details of the analysis. As discussed above the superpotential
 we consider has three sectors
\begin{equation}
W=W_V+W_H+W_{VH}
\end{equation}
where $W_V$ is the superpotential
in the visible sector which contains quarks, leptons and higgs fields,
$W_H$ is the superpotential in the hidden sector which contains
fields z whose VeV growth is of O($M_P$) and  leads to the breaking of
supersymmetry, and
$W_{HV}$ is the potential that contains exotic matter which couples to both
the fields of the visible sector and of the hidden
sector. For simplicity we assume that there are
no operators with dimensionality greater than 3 in W.
While the framework of the analysis we present is general and can be
implemented for any GUT group, we focus here on the extension of
the minimal SU(5) model. We assume
that the superpotential in the visible sector is given by
\begin{equation}
W_V=H_{1x}f_{1ij} K_{1ij}^x + H_2^xf_{2ij} J_{2xij}+(M_P s^2+(\lambda_1/3)
\Sigma^3+(\lambda_2/2)s \Sigma^2)+W_R
\end{equation}
 where i, j(=1,2,3) are the generation indices,
($H_{1x}$) is the $\bar5$-plet of Higgs, ($H_2^x$) is the 5-plet of Higgs,
($\Sigma$) is a 24-plet of Higgs
whose VeV growth
breaks SU(5)$\rightarrow$SU(3)xSU(2)xU(1), and s is a gauge singlet.
 f1,f2 are the Yukawa
couplings  and
 $J_{2ij}$ and  $K_{1ij}$ are the currents defined in eqs(3) and (4)
in terms of the three
generations of $\bar5$-plet of matter($M_{xi}^{'}$,i=1,2,3), and 10-plet
 of matter($M_i^{xy}$).
$W_R$ is the remaining part which gives masses to the Higgs triplets
while keeping the Higgs doublet light.
For exotic matter we make the
simple assumption that it belongs to $\bar 5+5$ representations
of SU(5) ( with equal number of $\bar 5$ and 5 for anomaly cancellation).
The sector which contains exotic matter, i.e. $W_{VH}$, has two
parts: the first part contains the coupling of exotic matter with the
hidden
sector fields $z_I$ while the second part couples exotic matter to the
 fields of the visible sector. In general, one needs three different
types of exotic fields to preserve matter parity. We shall distinguish
these by the subscript $a$ ($a=1,2,3$). This is
required by the fact that there are
three different $\bar 5$ and 5 plet of currents that can be formed out of
the fields of the visible sector and thus one needs three different types
of 5 and $\bar 5 $ of exotic fields to couple to these. Thus the three
 $\bar 5 $ -plets of currents
of dimensionality two that can be formed out
of the fields that enter the visible sector are
\begin{equation}
J_{1x}\equiv H_{1y}\Sigma^y_x, J_{2xij}\equiv -(1/8)\epsilon_{x\mu\nu
yz}M_i^{\mu\nu}
M_j^{yz},J_{3xi}\equiv M'_{yi}\Sigma_x^y
\end{equation}
Similarly there are three different 5-plet of currents
 $K^x_a=(K^x_{1ij},K^x_2,K^x_{3i})$ which can be constructed out of the
fields appearing in the visible sector. These are
\begin{equation}
K_{1ij}^x\equiv M_{yi}' M_j^{xy},K_2^x\equiv \Sigma^x_yH_2^y,K_{3i}^x\equiv
M_i^{xy}H_{1y}
\end{equation}
One introduces the $\bar 5$ plet of exotic fields $E_{ax}$ and 5 plet of
exotic fields $F_a^x$ so that the interactions $E_{ax}K_a^x$ and $F_a^x J_{ax}$
are matter parity invariant. For generality we introduce n pairs of
such fields and  denote these by $E_{\alpha ax},F_{\alpha a}^x$($\alpha$ =1,2,
..,n) and  write $W_{VH}$ as follows
\begin{equation}
 W_{VH}= C_{I\alpha a\beta b}^{(1)}z_IE_{\alpha a x}F_{\beta b}^x
+\eta_{Ax}C_{ABy}^{(2)x}\xi_B^y
\end{equation}

where  $\eta_{Ax}$
and $\xi_B^x$  are defined  by
\begin{equation}
\eta_{Ax}=(E_{\alpha a x},J_{ax}), \xi^x_B=(F_{\alpha a}^x,K_{ax})
\end{equation}
Since $E_{\alpha ax},F_{\alpha a}^x$ have
dimensionality 1 while $J_{ax},K_{ax}$ have dimensionality 2,
the assumption that  $W_{VH}$ has no operators of dimensionality greater than
three then requires that
\begin{equation}
C_{ab}^{(2)}=0,
C_{\alpha a \beta by}^{(2)x}=C_{\alpha a \beta b}^{(0)}\Sigma_y^x
\end{equation}
where $C_{\alpha a \beta b}^{(0)}$ is field independent.

After spontaneous breaking of supersymmetry one gets a mass generation
in the heavy sector with a mass term of $E_{\alpha ax}M_{P\alpha a\beta b}
F^x_{\beta b}$,
where $M_{P\alpha a\beta b}\equiv <z_I>C_{I\alpha a\beta b}$ and
$M_{P\alpha a\beta b}\sim O(M_P)$.
We proceed now to eliminate the superheavy fields $E_{\mu x},F_{\mu}^x$ (where
$\mu=(\alpha a)$) and
at the same time carry out an expansion in $\Sigma/M_P$. Thus, for example,
\begin{equation}
E_{\mu x}=-J_{ax}C_{a\nu}^{(2)}M^{-1}_{P\nu\mu}+J_{ay}C_{a\nu}^{(2)}
M^{-1}_{P\nu\rho}
C^{(0)}_{\rho\lambda}\Sigma^y_x M^{-1}_{P\lambda\mu}+ ...
\end{equation}
and a similar equation holds for $F^x_{\mu}$. Eliminating the heavy fields
in eq(5) and
carrying out a rediagonalization and
assuming for simplicity that,
$M_{P\alpha a\beta b}$=$M_P$$\delta_{\alpha\beta}\delta_{ab}$,
then gives an effective superpotential at the GUT scale of the form
\begin{eqnarray}
 W_{eff} &=&H_{1x}(U_{1ij}{\Sigma\over M_P}+U_{2ij} {\Sigma^2\over M_P^2}
+U_{3ij}{\Sigma^3\over M_P^3}+...)^x_yK^y_{1ij}\nonumber \\
& &\mbox{}+M'_{xi}(V_{1ji}{\Sigma\over M_P}+V_{2ji} {\Sigma^2\over
M_P^2}+V_{3ji} {\Sigma^3\over
M_P^3}+...)^x_yM^{yzj}H_{1z}\nonumber \\
& &\mbox{}+J_{2xij}(W_{1ij}{\Sigma\over M_P}+W_{2ij} {\Sigma^2\over
M_P^2}+W_{3ij} {\Sigma^3\over
M_P^3}+...)^x_yH_2^y
\end{eqnarray}

We find
that the effective interaction of quarks-leptons and Higgs at the GUT scale
contains a heirarchy of new mass scales via an expansion in ${\Sigma\over
M_P}$.This expansion arises because of the couplings of the exotic fields
with $\Sigma$ as can be seen from eqs(5-7).
 After spontaneous breaking of the GUT symmetry $\Sigma$ develops
a VeV so that $<\Sigma>=M(2,2,2,-3,-3)$ and the interactions of
the Higgs doublets and Higgs triplets at the GUT scale are given by

\begin{eqnarray}
W
&=&H_{1\alpha}l^{\alpha}_iA^E_{ji}e^c_j+H_{1\alpha}d^c_iA^D_{ji}q^{\alpha}_j+
H_2^{\alpha}u^c_iA^U_{ji}q_{aj}\nonumber \\
& &\mbox{}+(-M_{H3}H_{1a}H_2^a+ H_{1a}l_{\alpha
i}B^E_{ji}q^{\alpha}_{aj}+\epsilon_{abc} H_{1a}d^c_{bi}
B^D_{ji}u^c_{cj}\nonumber \\
& &\mbox{}+H^a_2 u^c_{ai}B^U_{ji}e^c_j +\epsilon_{abc}H_2^au_{bi}
C^U_{ji}d_{cj})
\end{eqnarray}

Here $A^{E,D,U}$, $B^{E,D,U}$, and $C^U$ are matrices in the generation space
and are determined in a power series
expansion in $\epsilon( \equiv M/M_P)$ using eq.(9). In the
analysis here we
 assume
that eq(2) contributes only to the up quark sector( even there only to the
top quark mass) and all of the remaining quark and lepton  masses arise from
higher dimensioned operators that are generated  when one  integrates
over the exotic fields. Thus we set $f_2$=0 in eq(2).
 The textures that enter in the $H_{1x}$ Higgs couplings
can be shown to satisfy the following sum rule
\begin{equation}
A^E+B^E+B^D=A^D
\end{equation}
Further as indicated  above these textures can be computed in a power series
expansion in terms of $\epsilon$.We get
\begin{eqnarray}
A^E &=& -3\epsilon
(U_1-V_1)+9\epsilon^2(U_2-V_2)-27\epsilon^3(U_3-V_3)+..\nonumber \\
A^D-A^E &=& -5\epsilon V_1+5 \epsilon^2 V_2-35 \epsilon^3
V_3+..\nonumber \\
B^E+A^E&=&-5\epsilon U_1+5 \epsilon^2 U_2-35 \epsilon^3 U_3+..
\end{eqnarray}
Similarly the textures that enter in the $H_2^x$ couplings can be
computed in a power series expansion in $\epsilon$.We get
\begin{equation}
A^U=f_1-3\epsilon W_1+9\epsilon^2 W_2+..\\
B^U-A^U=-5\epsilon(W_1-\epsilon W_2)+..,C^U=B^U\\
\end{equation}
There are several possible solutions[2-7] to the textures
$A^{E,D,U}$
at the GUT  scale which can lead to acceptable low energy quark-lepton
masses.We construct here one specific example.
It is seen that the choice[11].

\begin{eqnarray}
\epsilon U_{1ij} &=&-{1\over 3}D\delta_{i3}\delta_{j3},\nonumber \\
\epsilon^2U_{2ij} &=&{7\over15}E\delta_{i2}\delta_{j2}+{1\over 5}\delta
\delta_{i3}\delta_{j3},\nonumber \\
\epsilon^3U_{3ij} &=&-{1\over
27}F(\delta_{i1}\delta_{j2}+\delta_{i2}\delta_{j1} )\nonumber \\
f_{1ij} &=& A \delta_{i3}\delta_{j3},\nonumber\\
f_{2ij} &=& 0,\nonumber\\
-3\epsilon W_{1ij} &=& B
(\delta_{i2}\delta_{j3}+\delta_{i3}\delta_{j2}),\nonumber \\
9\epsilon^2W_{2ij} &=& C(\delta_{i1}\delta_{j2}+\delta_{i2}\delta_{j1}
)\nonumber\\
V_{1ij} &=& 0=V_{3ij},\nonumber\\
\epsilon^2V_{2ij} &=& {4\over 5}E\delta_{i2}\delta_{j2}
+ {1\over 5} \delta \delta_{i3}\delta_{j3}
\end{eqnarray}

 leads to

\begin{equation}
A^E=\left(\matrix{0&F&0\cr
                  F&-3E&0\cr
                  0&0&D\cr}\right),
A^D=\left(\matrix{0&Fe^{i\phi}&0\cr
                  Fe^{-i\phi}&E&0\cr
                  0&0&D+\delta\cr}\right),
A^U=\left(\matrix{0&C&0\cr
                  C&0&B\cr
                  0&B&A\cr}\right)
\end{equation}

which are the Georgi-Jarlskog textures when $\delta =0$[2,12].
{}From eq(14) we find that in the down quark
and lepton sectors one has
$D\sim O(\epsilon),E\sim O(\epsilon^2)$ and $F\sim
O(\epsilon^3)$, and in the up-quark sector one has $A\sim O(1)$,
$B\sim O(\epsilon)$ and
$C\sim O(\epsilon^2)$. These are the orders needed to correctly generate
the mass hierarchies in the down-quark and lepton
sectors and in the up-quark sector.
We turn now to a determination of the ratio $\epsilon$. Minimization of the
potential with respect to the fields s and $\Sigma$ determines $\epsilon
=-\lambda_1/(15 \lambda_2^2)$[13]. Thus, for example, with $\lambda_1\approx
-0.6$
 and $\lambda_2\approx 1.5$ one gets $\epsilon \approx 1/60$. With this value
of $\epsilon$  one
finds that the quark-lepton mass hierarchies are well reproduced with
the (remaining) Yukawa couplings O(1). Thus with $M_P=2.4\times 10^{18}$ GeV
implies an $M_G\sim 4\times 10^{16}$ GeV which is about the correct GUT scale.
We note that it is the $J_3...K_3$ interaction structure  which gives the
$\epsilon^2 V_2$ term in eq(14) and is responsible for breaking
the down-quark and lepton mass degeneracy.

Remarkably, the inputs of eq(14) which determine the quark-lepton textures
of eq(15) can also be used to compute the textures in the Higgs triplet
sector.
 We find

\begin{equation}
B^E=\left(\matrix{0&(-{19\over 27}+e^{-i\phi})F&0\cr
                  (-{19\over 27}+e^{i\phi})F&{16\over 3}E&0\cr
                  0&0&({2\over 3}D+\delta)\cr}\right)
\end{equation}
\begin{equation}
B^D=\left(\matrix{0&-{8\over 27}F&0\cr
                  {-8\over 27}F&-{4\over 3}E&0\cr
                  0&0&-{2\over 3}D\cr}\right),
B^U=\left(\matrix{0&{4\over 9}C&0\cr
                  {4\over 9}C&0&-{2\over 3}B\cr
                  0&-{2\over 3}B&A\cr}\right)
\end{equation}
Comparison of eqs. (15) and (16,17) shows that the textures in the Higgs
triplet sector are different than those in the Higgs doublet sector.
 This fact
has important new consequences for the predictions of the proton decay modes
as we now discuss. The  analysis of the above problem requires that one use
renormalization group equations to scale down the textures in the quark-lepton
sector from the GUT scale to the electro-weak scale. One then goes to the
basis where the quark-
lepton mass matrices are diagonal by defining transformations

\begin{equation}
U^{\dagger}_RA^UU_L=A^{U(d)},\\
D^{\dagger}_RA^DD_L=A^{D(d)},\\
E^T_LA^EE_R^*=A^{E(d)}
\end{equation}
Next we  turn  to the baryon-number violating interactions in this basis.
After integrating over the heavy Higgs triplet fields the baryon number
violating dim5 operator is given by

\begin{eqnarray}
W_5 &=& {1\over M_{H3}}(-e_LU^{(1)}B^{E(d)}V^{(1)\dag}u_{La}\nonumber \\
&+&\mbox{}\nu U^{(1)}B^{E(d)} U^{(2)}) \epsilon_{abc}u_{bL}PU^{(2)}
B^{U(d)}V^{(2)}d_{cL} \nonumber \\
&+&\mbox{}\left({1\over M_{H3}}\right) (u^c_a A^{U(d)}V^{(3)}e^c
\epsilon_{abc}d^c_b U^{(3)}B^{D(d)} V^{(4)\dag}P^*u^c_c)
\end{eqnarray}

In the above the first two sets of terms are the LLLL type dim5 operators,
and the last term  is the RRRR type dim5 operator, $P\equiv U_L^TU_R$,and
        $U^{(i)}(i=1,2,3)$ are defined by

\begin{equation}
U^{(1)}=E^T_LE^*_{1L}, U^{(2)}=U^{\dag}_RU_{1R},U^{(3)}=D_R^{\dag}D_{1R}
\end{equation}

where $B^E$,$B^U$ and $B^D$ are diagonalized by the transformations

\begin{equation}
E^T_{1L}B^EE_{1R}^*=B^{E(d)},U_{1R}^{\dag}B^UU_{1L}=B^{U(d)},
D_{1R}^{\dag}B^DD_{1L}=B^{D(d)}
\end{equation}

The matrices $V^{(i)}$ are defined so that

\begin{equation}
V^{(1)}\equiv U_L^{\dag}E_{1R}^* ,V^{(2)}\equiv U_{1L}^{\dag}D_L,
V^{(3)}\equiv U_L^{\dag}E_{R}^*, V^{(4)}\equiv U_L^{\dag}D_{1L}
\end{equation}

There have been extensive previous analyses of proton stability[14-15]
 in the literature. All of these [14,15] have ignored the
effect of the textures on p-decay lifetime.
 In the usual analyses one uses the
assumption that
$B^E=A^E,B^D=A^D$ and $B^U=A^U$.
In this limit $U^{(i)}(i=1,2,3)$ are all unity, and $V^{(i)}(i=1..4)$
equal the KM matrix $V_{KM}$.
 It has recently been realised
that quark-lepton textures can affect p-lifetimes.A generic discussions
of this effect is given in references[16,17].
 In this Letter we have given a full solution to the
problem of how proton decay interactions depend on the textures in the
context of a specific model.
 These textures
are model dependent. Thus different assumptions about the $A^{E,D,U}$
textures at the GUT scale would in general lead to different $B^{E,D,U}$
textures in the Higgs triplet sector.
One of the consequences of the above analysis is that one cannot diagonalise
$A^E$ and $B^E$ by the
same tranformations. A similar situation holds for $A^D$ and $B^D$ and for
$A^U$ and $B^U$.
Because of this there are  modifications of the predictions
of p-decay lifetime. A full numerical analysis of proton decay lifetimes in the
context of the new textures is outside the scope of this paper. However, one
can make a rough estimate of the size of the change that one expects.
For the most dominant  $p-decay$ mode $p\rightarrow \bar\nu K$ we note that if
we ignore the effect of
the mismatch matrices $U^{(i)}$ and assume that $V^{(i)}$ are all the same,
 then the change in the lifetime of this
 decay mode can be estimated as follows: for the standard case discussed
previously the $p-$ decay width
is proportional to $(m_c m_s)^2$.With the new textures
the decay width will be proportional to $((4/9)m_c(16/9)m_{\mu})^2$ as can be
seen by a comparison on the eigenvalues of A and B.This would imply
an enhancement in the p-decay lifetime for this mode  by a factor of
about 3. We also note that the
appearance of the CP violating phase in the GUT sector as given by
eq (16) may have implications for cosmology such as in the generation of
baryonic asymmetry in the early universe.

\section*{Acknowledgements}
This research was supported in part by NSF grant number PHY--19306906 and
at the Institute for Theoretical Physics in Santa Barbara under grant number
PHY94-07194.


\begin{thebibliography}{99}
\bibitem{ref11} A  satisfaction of eq(14) in terms of $C_{AB}^{(2)}$
of eq(5) is straightforward.
\bibitem{reftex} The textures of eq(15) with $\delta=0$  lead to an
exact b-$\tau$ unificaion at the GUT scale.
One can generate a b-$\tau$ splitting by allowing a non-vanishing value
for $\delta$.
Since $\delta/D$ is O($\epsilon$), one can
expect a few percent correction to the exact $b-\tau$ equality at the
GUT scale.
\bibitem{ref12} In terms of $\epsilon$ the mass of the heavy X vector boson
associated with the breaking of the SU(5) symmetry is given by $M_V/M_P=
5\sqrt 2 g_5 \epsilon$, and the Higgs triplet mass is given by $M_{H3}/M_P
=5\lambda_2\epsilon$.
\end{thebibliography}
\end{document}